\documentclass[manuscript=article, layout=twocolumn]{achemso}

\def\by#1{#1,}
\def\and{and }
\def\yr#1{{\bf #1}}
\def\paper#1{#1}
\def\jour#1{{\it #1}}
\def\vol#1{{\it #1},}
\def\pages#1{\hbox{#1},}
\usepackage{amsmath}
\usepackage{amsfonts}
\usepackage{amssymb}
\usepackage{graphicx}
\usepackage{appendix}
\usepackage{placeins}
\usepackage{achemso}

\title{Hydrodynamics of moving contact lines: macroscopic versus microscopic.}
\author{Alex V. Lukyanov}
\email{a.lukyanov@reading.ac.uk}
\author{Tristan Pryer}
\affiliation{School of Mathematical and Physical Sciences, University of Reading, Reading RG6 6AX, UK}

\begin{document}

\begin{abstract}
The fluid-mechanics community is currently divided in assessing the boundaries of applicability of the macroscopic approach to fluid mechanical problems. Can the dynamics of nano-droplets be described by the same macroscopic equations as the ones used for macro-droplets? To the greatest degree, this question should be addressed to the moving contact-line problem. The problem is naturally multiscale, where even using a slip boundary condition results in spurious numerical solutions and transcendental stagnation regions in modelling in the vicinity of the contact line. In this publication, it has been demonstrated {\it via} the mutual comparison between macroscopic modelling and molecular dynamics simulations that a small, albeit natural, change in the boundary conditions is all that is necessary to completely regularize the problem and eliminate these nonphysical effects.  The limits of macroscopic approach applied to the moving contact-line problem have been tested and validated from the first microscopic principles of molecular dynamic simulations.         
\end{abstract}

\medskip\noindent
Keywords: {\it wetting, nano-scale, contact line, macroscopic boundary conditions, molecular dynamics simulations.}
\medskip

\section*{Introduction}
The modelling of the wetting of a solid substrate by a liquid on the continuum scale is a general problem in science and in the emerging industrial applications of nanofluidics~\cite{Beebe2002, Quake2005, Rauscher2008, Ralston2008, Andreotti2013}. This problem often involves moving contact lines and requires such a set of macroscopic boundary conditions for the Navier-Stokes equations to ensure the macroscopic problem is well-posed and also demonstrates the correct kinematics of the simulated flows. It is well known that the standard no-slip boundary condition in a moving contact-line problem leads to a non-integrable stress singularity at the contact line such that no solution to the whole hydrodynamic problem exists~\cite{Huh-Scriven1971}. Removal of this singularity can be achieved through the introduction of finite slip into the macroscopic boundary conditions~\cite{Huh-Scriven1971, Dussan1976, Huh-Mason1977, Kirkinis2013}. Nevertheless, even if the no-slip condition is relaxed, the macroscopic problem is susceptible to spurious numerical solutions and, irrespective of the slip model, features nonphysical stagnation regions and weak singularities of pressure~\cite{Shikhmurzaev2006, Shikhmurzaev-2007, Sprittles2011}, which are not observed in experiments~\cite{Dussan1974, Clarke1995, Garoff1997, Breuer2015}. An attempt to remove artificial features in capillary flow modelling has been made in the interface formation theory (IFT), where the liquid motion is rolling and the pressure is regular~\cite{Shikhmurzaev2006, Shikhmurzaev-2007, Sprittles2013}. 

However, the IFT required an essential assumption of macroscopically finite, rather long relaxation times of the surface phases, which was found not to be the case in monatomic and bead-spring Lennard-Jones (LJ) model fluids {\it via} direct molecular dynamics simulations (MDS)~\cite{Blake2009, Lukyanov2013-1, Lukyanov2013-2}. The short, practically microscopic relaxation time of the surface phase obtained in the MDS was mostly conditioned by the characteristic intrinsic width of the interfacial region~\cite{Lukyanov2013-1, Lukyanov2013-2}, which, as it turned out, was about one atomic diameter or smaller even for water liquid-gas interfaces~\cite{Ventikos2004, Grest2006}. The MDS results were supported by experimental evidence. The intrinsic width obtained in the MDS of monatomic LJ model fluids (when the model LJ interaction potential of the particles was very close to that measured experimentally, for example between argon atoms) was found to be in very good agreement with that estimated from experiments~\cite{Beaglehole1979, Beaglehole1980, Ventikos2004, Penfold2001}. Similar values of the intrinsic width were reported for the liquid-gas interfaces of carbon tetrachloride, water, alkanes and alcohols~\cite{Penfold2001}. Therefore, it is difficult at the moment to expect, at least for simple interfaces, that the relaxation times of the surface phase would be on the time scale required by the IFT. Obviously,  another solution within the macroscopic framework should be found to remove artificial features in the modelling at the contact line.

While the presence of a stagnation point seems to be insignificant for simulations of large scale flows, consider for example the asymptotic analysis by Cox~\cite{Cox1986}, the stagnation zone can substantially affect the kinematics of nano-flows and simulation of capillary flows with complex interfaces laden with surfactant molecules or nanoparticles~\cite{Deegan1997, Beebe2005, Vermant2012}. In the latter, the stagnation zone will significantly impede the motion of surfactant molecules or nanoparticles creating either artificial surface tension gradients in the macroscopic solutions or a conglomeration of nanoparticles, which in turn may substantially affect a simulation of particle assembly processes~\cite{KineticAssembly2006, Yunker2011}. At the same time, the estimated slip lengths for a variety of liquid-solid combinations lie in the mesoscopic range~\cite{Lauga2005, Yoda2010}, where the notion of the stress tensor, the main quantity in macroscopic description, is ill-defined~\cite{Schofield1982, Rowlinson-Widom-1989}. The question then appears to what extent and how can we model such flows using macroscopic approach? To answer these questions we will turn to MDS of bead-spring model fluids and directly compare the MDS results with macroscopic modelling using a modified set of boundary conditions naturally derived from the microscopic principles. It is demonstrated that the problem can be fully adequate without resorting to complex mesoscopic approaches, such as diffuse interface models~\cite{{Seppecher1996, Qian2003}}, even if the surface phase relaxation time is macroscopically zero and there are mesoscopic length scales involved.

\begin{figure}
\begin{center}
\includegraphics[trim=-1.5cm 3.cm 1cm -0.5cm,width=\columnwidth]{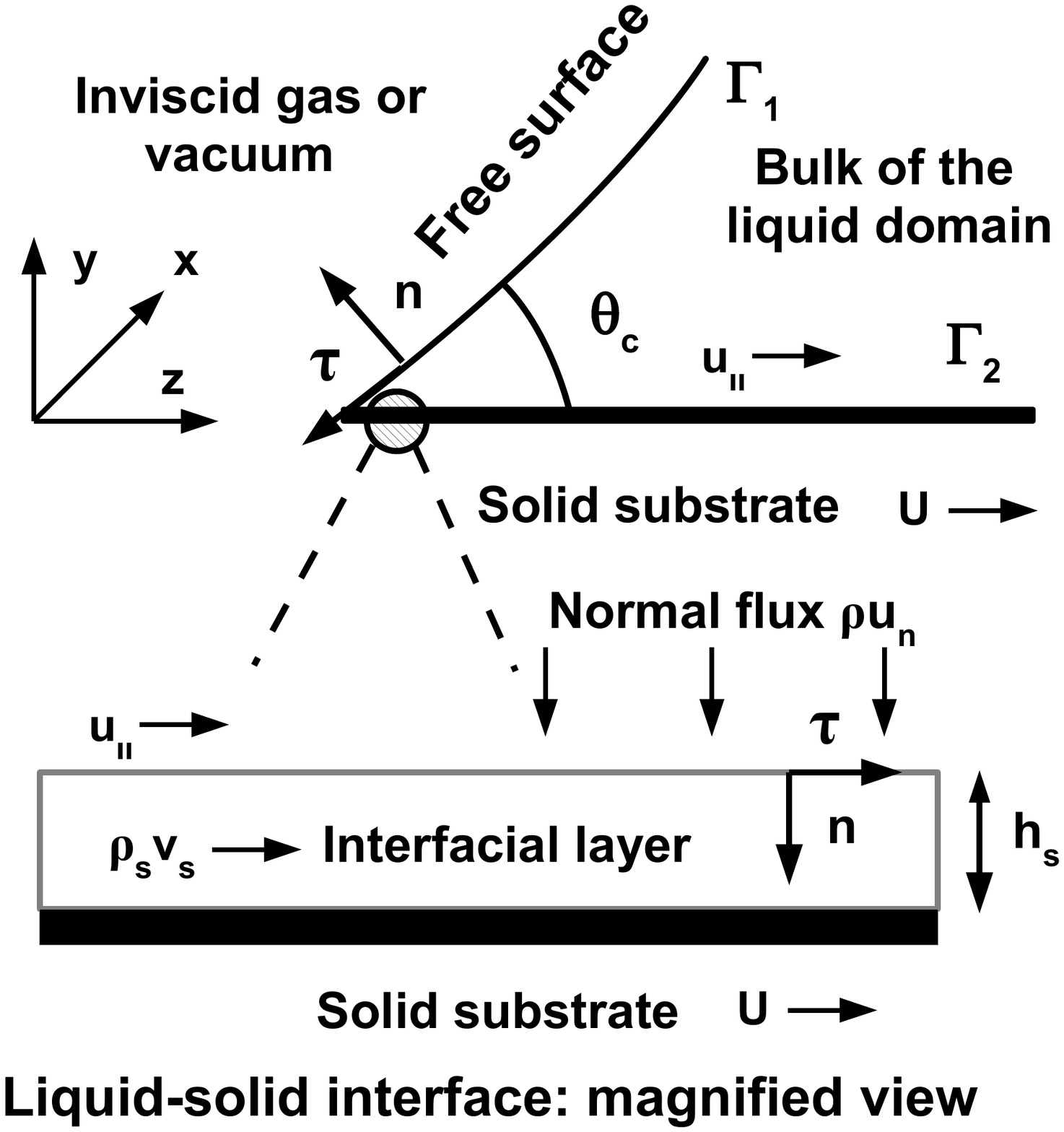}
\end{center}
\caption{An illustration of the moving contact line problem in two-dimensional geometry with dynamic contact angle $\theta_c$ and the magnified view of the liquid-solid boundary layer. The free surface is designated by $\Gamma_1$ and the liquid-solid interface is designated by $\Gamma_2$. The contact line is at rest and the substrate is moving in the tangential direction with velocity  ${\bf U}=(0,0,U)$.} 
\label{Fig1}
\end{figure}

In this study, we focus on the steady motion of a contact line of a Newtonian liquid in a two-dimensional case in an inviscid gas or vacuum over a stationary homogeneous and flat substrate, Figure \ref{Fig1}. All liquid interfaces are assumed to be simple, that is not laden with surfactant molecules and/or nanoparticles, and in isothermal conditions. The later approximation should hold even in extreme wetting conditions. Indeed, assuming no singular sources of the heat production (no singular shear rates) in the contact line region with a characteristic length scale $L_0$  and a steady state, balancing the thermal energy flux from the contact line region with the viscous dissipation in it, one has  $\Delta T_0 \approx  \mu \dot{\gamma}^2 L_0^{2} \kappa^{-1}$. Here, $\Delta T_0$ is the temperature variation with respect to a bulk value, $\mu$ is liquid viscosity, $\kappa$ is the coefficient of thermal conductivity of the liquid, $\dot{\gamma}$ is the shear rate. Then for water at room temperature, $L_0=1\,\mbox{nm}$ and $\dot{\gamma}=10^{10}\,\mbox{s}^{-1}$, $\Delta T_0\approx 0.2\,\mbox{K}$. So that, large temperature variations or gradients at the contact line, which may potentially create surface tension gradients leading to strong Marangoni effects, are only expected at hydrodynamic velocities approaching the liquid thermal velocity, that is at much higher shear rates $\dot{\gamma}\approx 10^{11}\,\mbox{s}^{-1}$. 

In the reference frame moving with the contact line, the dynamic contact angle $\theta_c$ is assumed to be a function of the substrate velocity $U$ according to the modified Young-Dupr\'{e} equation~\cite{Blake1969}, $\gamma\cos\theta_c =\gamma_{LS}-\gamma_{GS} -F(U)$, where $\gamma$, $\gamma_{GS}$ and $\gamma_{LS}$ are the liquid-gas, gas-solid and liquid-solid coefficients of surface tension respectively, and $F(U)$ is the velocity dependent friction force acting on the contact line per unit length~\cite{Nano2016}. The origin of that singular force $F$ has been studied in detail using MDS~\cite{Nano2016}. It has been shown that the force is the consequence of the microscopic processes taking place at the contact line on the length scale of a few atomic distances, which is induced by the interaction potential of the constituent molecules. The observed length scale defines the size of the contact line zone. The result of the microscopic interactions is nonlinear friction force distribution acting on the first monolayer at the solid substrate. The integral of the distribution over the contact line zone results in the singular force $F$, which manifests in the modified Young-Dupr\'{e} equation.  In this work we consider macroscopic interaction of the bulk liquid with its interfaces beyond the contact line zone, that is the boundary conditions to the Navier-Stokes equations.     

\section*{Models and simulation methods}
Experimental studies of dynamic wetting phenomena have been focusing on the behaviour of the apparent dynamic contact angle as a function of the contact line velocity, while much lesser attention has been paid to velocity distributions in the proximity of the contact line~\cite{Ralston2008}. In part, this was caused by the experimental difficulties with the spatial resolution of the velocity field at the contact line region. Earlier studies of the velocity distribution at the contact line by means of particle tracking velocimetry had spatial resolution up to several micrometres at best, which was nevertheless sufficient to verify available asymptotic theories of the flow motion at the contact line~\cite{Clarke1995, Garoff1997}. Recently, spatial resolution about $50\,\mbox{nm}$ has been achieved using the total internal reflection fluorescence technique~\cite{Breuer2015}. While this was the great achievement, it was still insufficient to resolve velocity in the proximity of the contact line region given experimentally observed slip lengths often found around ten nanometers or much less~\cite{Lauga2005, Yoda2010}. At the same time, it is in that region where the macroscopic approximation may break down and is in need for corrections. To verify a set of macroscopic boundary conditions formulated to correct macroscopic problem descriptions at the contact lines, we turn to MDS where the spatial resolution at the contact line is not an issue. In particular, we consider a problem of a meniscus forced to move at constant velocity in between two solid substrates separated by the distance of tens of nanometers, which is a simplified two-dimensional version of a common experimental set-up with a meniscus moving in a hollow needle under elevated pressure conditions.   

\subsection*{Macroscopic boundary conditions}
To understand how the boundary conditions should be modified at the contact line, consider first the Navier slip condition~\cite{Huh-Scriven1971, Dussan1976}, which is simply a linear relationship between the variation of the tangential velocity across an interfacial layer and the tangential stress acting from the bulk liquid on the interface, 
$$
\mu\, {\bf n} \cdot (\nabla {\bf u}+(\nabla {\bf u})^{T})\cdot {\bf T} \left. \right|_{\Gamma_2}
$$ 
\begin{equation} 
\label{NSOr}
=\left. \beta ({\bf U-u})\cdot {\bf T} \right|_{\Gamma_2},  
\end{equation}
see Figure \ref{Fig1} for an illustration.
Here, ${\bf u}$ is the hydrodynamic velocity in the bulk, ${\bf n}$ is the outward pointing normal vector, the tensor ${\bf T= I-nn}$  extracts the tangential to the interface component of a vector, $\mu$ is liquid viscosity and $\beta$ is the coefficient of sliding friction. The ratio $\mu/\beta$ has the dimension of length and represents the apparent slip length $L_s=\mu/\beta$. The application of the boundary condition (\ref{NSOr}) along with the impermeability condition ${\bf u\cdot n}=0$ leads to a stagnation zone at the contact line and a logarithmic singularity in the pressure, such that the radial velocity behaves as $u_{r}\propto r$ and the pressure $p\propto\ln r$ when approaching the contact line region $r\to 0$~\cite{Shikhmurzaev2006, Shikhmurzaev-2007}.

Within the framework of non-equilibrium thermodynamics, the boundary condition (\ref{NSOr}) appears as a linear phenomenological law between fluxes and thermodynamic forces occurring in the singular entropy production in the liquid-solid interfacial layer~\cite{Bedeaux1976}. In this approach the interfaces can carry singular particle density $\rho_s$, which is not an excess quantity, as it would be in the Gibbs' formulation, but the total number of liquid particles in the interface per unit area transported with surface velocity ${\bf v}_s$~\cite{Shikhmurzaev2006, Shikhmurzaev-2007}. This method of treatment developed in the works of Bakker, Guggenheim and in the IFT, ~\cite{Rowlinson-Widom-1989, Shikhmurzaev-2007} and references therein, is more convenient for formulation of boundary conditions for macroscopic equations, such as the Navier-Stokes equations. 

In the formulation, $\rho_s$ and $\gamma$ are surface scalars and ${\bf v}_s$ is a surface vector. That is, if considering them as functions of all three spatial coordinates defined on a smooth surface $\Gamma$ with a normal vector $\bf n$, for example, $({\bf n}\cdot \nabla) {\bf v}_s=0$. The divergence of a surface vector then is $\nabla\cdot {\bf v}_s=\nabla\cdot ({\bf v}_s \cdot {\bf T})+({\bf v}_s\cdot {\bf n})\, \nabla\cdot {\bf n}$, where $\nabla\cdot ({\bf v}_s \cdot {\bf T})$ can be calculated using a parametrization of the surface $\Gamma$. In a steady state, when ${\bf v}_s\cdot {\bf n}=0$, $\nabla\cdot {\bf v}_s=\nabla\cdot ({\bf v}_s \cdot {\bf T})$.

Therefore, any changes in the tangential velocity ${\bf u\cdot T}\left. \right|_{\Gamma_2}$ at the interface inevitably result in a change of the total flux in the interfacial layer $\rho_s^{(2)} {\bf v}_s\cdot {\bf T}\left. \right|_{\Gamma_2}$, Figure \ref{Fig1}. Further, we will often use the standard shorthand notation $\rho_s^{(i)}$ for $\rho_s$ on $\Gamma_i$, where the superscripts $(1)$ and $(2)$ indicate the liquid-gas and liquid-solid interfaces respectively. Then, simply by conservation of mass in the interfacial layer one has  
\begin{equation}
\label{Conservation-Mass}
\frac{\partial \rho_s^{(2)}}{\partial t}+\nabla\cdot  (\rho_s^{(2)} {\bf v}_s)\left.  \right|_{\Gamma_2}=\left. \rho {\bf u\cdot n}\right|_{\Gamma_2}
\end{equation}
where $\rho$ is the bulk liquid density. Based on the previous study of the surface phase relaxation time in bead-spring model fluids~\cite{Lukyanov2013-2}, we may presume that it is negligible on the macroscopic time scale of hydrodynamic motion at temperatures far from the critical point of the liquid. Hence, for isothermal flows and homogeneous substrates the liquid solid interface can be defined by $\rho_s^{(2)}=const$. Then, from (\ref{Conservation-Mass})
\begin{equation}
\left. \rho\, {\bf u\cdot n}\right|_{\Gamma_2}=\rho_s^{(2)} \nabla\cdot {\bf v}_s\left. \right|_{\Gamma_2}.
\label{MImpC}
\end{equation}
So, the introduction of finite slip in the boundary conditions implies that the standard impermeability condition should be also relaxed at the contact line. At a stationary free surface, similar to (\ref{MImpC}) assuming $\rho_s^{(1)}=const$
\begin{equation}
\left. \rho\, {\bf u\cdot n}\right|_{\Gamma_1}= \rho_s^{(1)} \nabla\cdot {\bf v}_s\left. \right|_{\Gamma_1}.
\label{MImpCf}
\end{equation}
Note that both boundary conditions (\ref{MImpC}) and (\ref{MImpCf}) can be obtained within the IFT framework in the limit of zero relaxation time and negligible surface tension gradients $\nabla\gamma=0$~\cite{Shikhmurzaev2006, Shikhmurzaev-2007}. 

The boundary conditions (\ref{MImpC}) and (\ref{MImpCf}) should be complemented with the conservation of surface flux at the contact line  
\begin{equation}
 \rho_s^{(1)} {\bf v}_s\cdot {\bf \tau} \left. \right|_{\Gamma_1} =  \rho_s^{(2)} {\bf v}_s\cdot {\bf \tau}\left.  \right|_{\Gamma_2}
\label{CLflux}
\end{equation} and the functional dependencies  
\begin{equation}
\label{Free-surface-no-slip}
\left. {\bf v}_s\cdot {\bf T} \right|_{\Gamma_1}=\left. {\bf u}\cdot {\bf T} \right|_{\Gamma_1}
\end{equation} 
\begin{equation}
\displaystyle \left. {\bf v}_s\cdot {\bf T} \right|_{\Gamma_2}=  \frac{{\bf u}+{\bf U}}{2}\cdot \left. {\bf T} \right|_{\Gamma_2}
\label{Coutte} 
\end{equation} 
to eliminate surface velocity ${\bf v}_s$ from the model. Here $\bf \tau$ is the unit tangential vector to the interface, which is normal to the contact line, Figure \ref{Fig1}. The functional relationships (\ref{Free-surface-no-slip}) and (\ref{Coutte}), as is seen, can be interpreted as the plug and Couette flows in the boundary layers respectively, and can be obtained in the same limit of zero relaxation time in the IFT~\cite{Shikhmurzaev2006, Shikhmurzaev-2007}. Indeed, in the IFT in general, in the framework of non-equilibrium thermodynamics, on a free surface
\begin{equation}
\label{IFT-1}
\left. 4 B ({\bf v}_s - {\bf u} )\cdot {\bf T} \right|_{\Gamma_1}=\left.(1+4 A B)\nabla \gamma\right|_{\Gamma_1}
\end{equation}
and on a solid substrate
\begin{equation}
\label{IFT-2}
\displaystyle \left. {\bf v}_s\cdot {\bf T} \right|_{\Gamma_2}=  \frac{{\bf u}+{\bf U}}{2}\cdot \left. {\bf T} \right|_{\Gamma_2} + \left. C \nabla \gamma\right|_{\Gamma_2},
\end{equation}
where $A$, $B$ and $C$ are some constant phenomenological parameters of the theory~\cite{Shikhmurzaev2006, Shikhmurzaev-2007}. Then, in the absence of any surface active molecules, that is in the approximation of simple interfaces, and in isothermal conditions, that is in the absence of any gradients of temperature at the interfaces, surface tension is at equilibrium $\nabla \gamma=0$ in the limit of zero relaxation time, and the results (\ref{Free-surface-no-slip}) and (\ref{Coutte}) follow from (\ref{IFT-1}) and (\ref{IFT-2}).

At the free surface one has also the condition of zero tangential force acting on the interface 
\begin{equation}
\left. \mu\, {\bf n} \cdot (\nabla {\bf u}+(\nabla {\bf u})^{T})\cdot {\bf T} \right|_{\Gamma_1}=0
\label{tangential-stress}
\end{equation} 
and the dynamic condition (at zero external pressure)
$$
\left \{-p + \mu\, {\bf n} \cdot (\nabla {\bf u}+(\nabla {\bf u})^{T})\cdot {\bf n}\right \}\left. \right|_{\Gamma_1}
$$ 
\begin{equation}
=-\gamma\nabla\cdot {\bf n}\left. \right|_{\Gamma_1},
\label{normal-stress}
\end{equation} 
where $p$ is pressure in the liquid. Choosing $L_s$, $U$ and $\frac{\mu U}{L_s}$ as the characteristic length, velocity and pressure respectively, the system of governing equations for the moving contact line problem in the limit of $Re=\frac{\rho U L_s}{\mu}\ll 1$ takes the standard form of the Stokes equations
\begin{equation}
\label{Stokes}
\nabla\cdot {\bf u}=0,\quad \nabla p=\Delta {\bf u}.
\end{equation}
The boundary conditions (\ref{NSOr}), (\ref{MImpC})-(\ref{normal-stress}) after eliminating ${\bf v}_s$ become
at the liquid-solid interface
\begin{equation}
{\bf u\cdot n}\left. \right|_{\Gamma_2} =\alpha_2\, ({\bf T}\cdot \nabla) \cdot {\bf u}\left. \right|_{\Gamma_2},
\label{MImpCND}
\end{equation} 
\begin{equation}
 {\bf n} \cdot (\nabla {\bf u}+(\nabla {\bf u})^{T})\cdot {\bf T} \left. \right|_{\Gamma_2}=({\bf \bar{U}-u})\cdot {\bf T} \left. \right|_{\Gamma_2},
\label{GNavierND}
\end{equation}
at the free surface
\begin{equation}
{\bf u\cdot n}\left. \right|_{\Gamma_1}=\alpha_1 \, ({\bf T}\cdot \nabla) \cdot {\bf u}\left. \right|_{\Gamma_1}, 
\label{MImpCfND}
\end{equation}
\begin{equation}
{\bf n}\cdot (\nabla {\bf u}+(\nabla {\bf u})^{T}) \cdot {\bf T} \left. \right|_{\Gamma_1}=0
\label{tangential-stressND}
\end{equation}
$$
Ca\{-p + {\bf n}\cdot (\nabla {\bf u}+(\nabla {\bf u})^{T}) \cdot {\bf n}\}\left. \right|_{\Gamma_1}
$$ 
\begin{equation}
=-\nabla\cdot {\bf n}\left. \right|_{\Gamma_1}
\label{normal-stressND}
\end{equation}
and at the contact line
\begin{equation}
\alpha_1 {\bf u}\cdot {\bf \tau} \left. \right|_{\Gamma_1} =  \alpha_2 ({\bf u}+{\bf \bar{U}})\cdot {\bf \tau} \left. \right|_{\Gamma_2} 
\label{CLfluxND} 
\end{equation} 
with the non-dimensional parameters $\alpha_1=\tfrac{\rho_s^{(1)}}{\rho L_s}$, $\alpha_2=\tfrac{\rho_s^{(2)}}{2\rho L_s}$, $Ca=\tfrac{\mu U}{\gamma}$ (the capillary number) and the velocity vector ${\bf \bar{U}}=(0,0,1)$.

\begin{figure}
\begin{center}
\includegraphics[trim=0cm 5.5cm 0cm 0cm,width=\columnwidth]{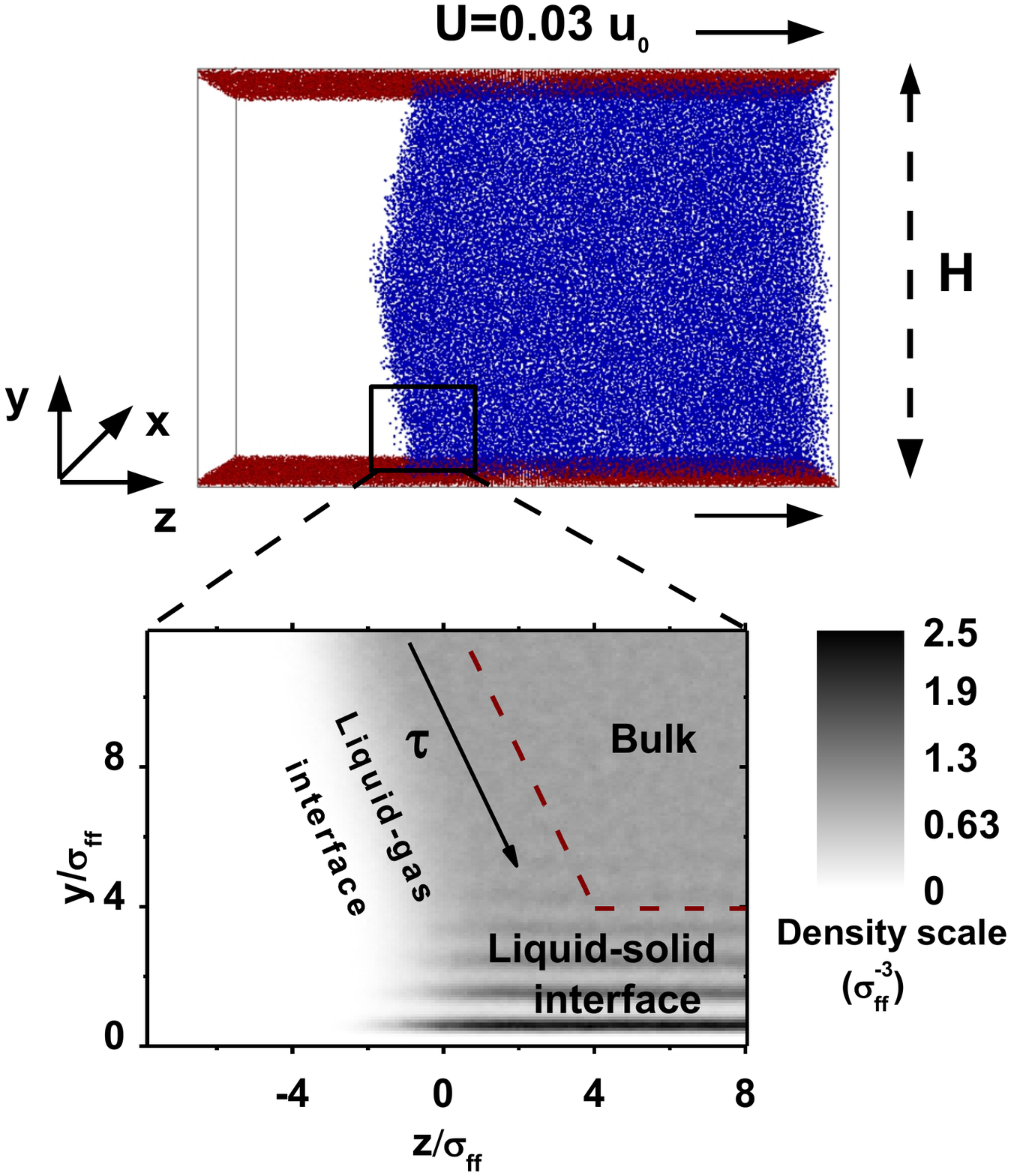}
\end{center}
\caption{Snapshot of the moving droplet simulation set-up in MDS at $U=0.03\, u_0$ and a dynamic contact angle $\theta_c=114\pm 4^{\circ}$, $u_0=\sqrt{\epsilon_{\mathit{ff}}/m_{\mathit{f}}}$. The static contact angle $\theta_0=39\pm 3^{\circ}$. The set-up is periodic in the $x$-direction with $H=60\,\sigma_{\mathit{ff}}$. The magnified view is the distribution of particle density. The dashed lines (red) designate the bulk-interface boundary. The distance $z$ is measured along the substrate from the equimolar point of the density distribution in the first monolayer $0\le y\le 1.1\,\sigma_{\mathit{ff}}$ and the distance $y$ is measured from the equimolar point of the solid particle distribution in the first layer facing the liquid in the lower substrate.} 
\label{Fig2a}
\end{figure}
\FloatBarrier

\subsection*{Molecular dynamics simulations}

\begin{figure}
\begin{center}
\includegraphics[trim=0cm 3.5cm 0cm 0cm,width=\columnwidth]{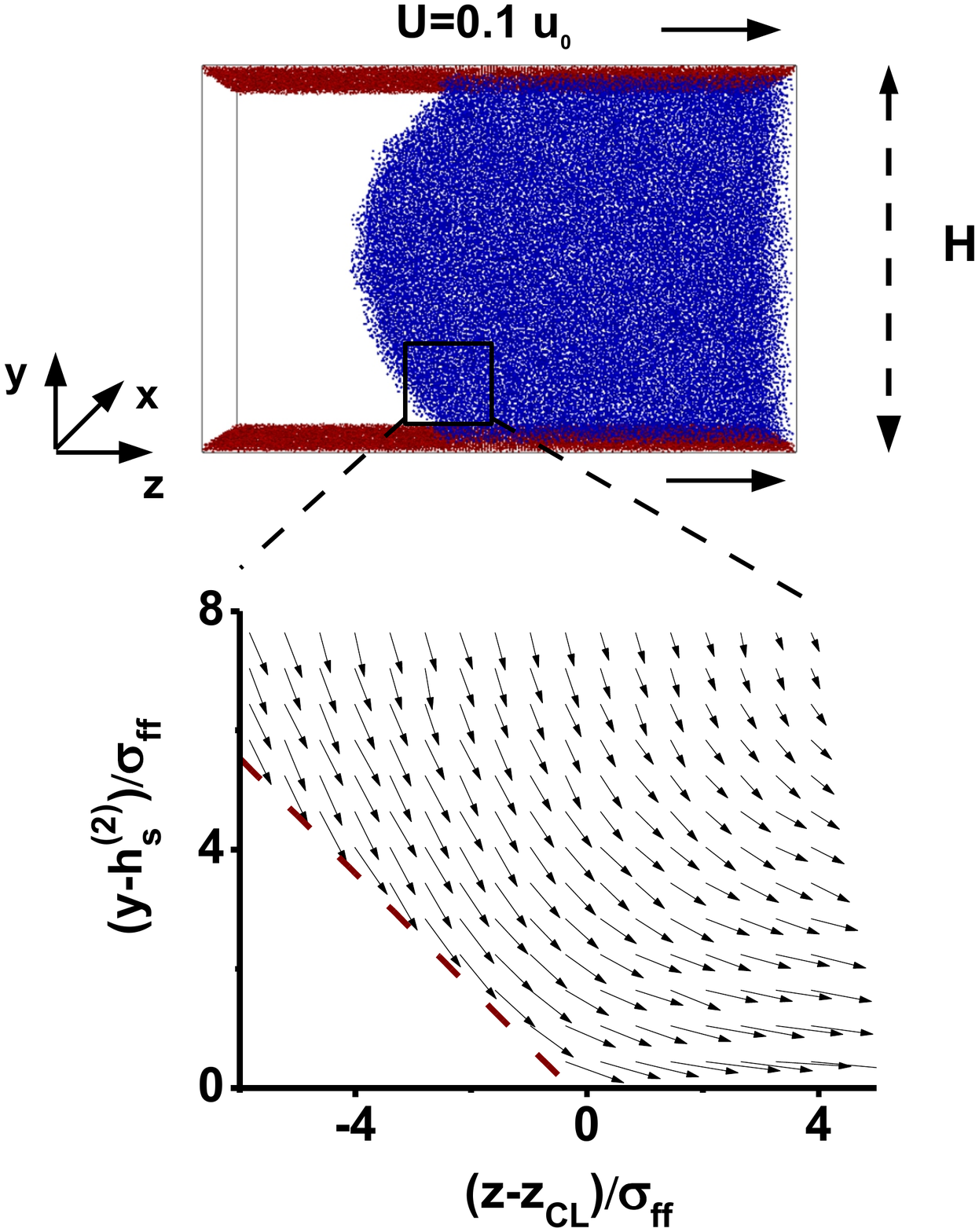}
\end{center}
\caption{Snapshot of the moving droplet simulation set-up in MDS at $U=0.1\, u_0$ and a dynamic contact angle $\theta_c=136\pm 5^{\circ}$, $u_0=\sqrt{\epsilon_{\mathit{ff}}/m_{\mathit{f}}}$. The static contact angle $\theta_0=39\pm 3^{\circ}$. The set-up is periodic in the $x$-direction with $H=60\,\sigma_{\mathit{ff}}$. The magnified view is the distribution of the flow field in the bulk. The size of the arrows is proportional to the velocity magnitude. The dashed lines (red) designate the bulk-interface boundary. In the plot, distances $y$ and $z$ are measured as in Figure \ref{Fig2a}.} 
\label{Fig2b}
\end{figure}

Before analyzing the macroscopic problem (\ref{Stokes})-(\ref{CLfluxND}), we examine the flow at the three-phase contact line region by means of MDS of a large, $60000$ particles of mass $m_f$, cylindrical liquid droplet forced to move with constant velocity ${U}$ between two identical solid substrates in periodic in the $x$-direction geometry, Figure \ref{Fig2a} and Figure \ref{Fig2b}. Each substrate is made of three face-centered cubic lattice layers of particles of mass $m_w=10 m_f$. Indexes $f$ and $w$ designate liquid and substrate particles respectively. 

The MDS setup is similar to that used in the studies of the contact line force~\cite{Nano2016}, so that here we only recollect the main aspects of the technique. Both substrate and liquid particles interact {\it via} the LJ potentials  $\Phi_{LJ}^{ij}(r)=4\epsilon_{ij} \left(\left(\frac{\sigma_{ij}}{r}\right)^{12}-\left(\frac{\sigma_{ij}}{r}\right)^{6}\right)$ with the cut off distance $2.5\,\sigma_{ij}$. Here $r$ is the distance between the particles, $\epsilon_{ij}$ and $\sigma_{ij}$ are characteristic energy and length scales. The liquid particles in most simulations performed in this study (unless otherwise stated) are combined into linear chains of $N_B=5$ beads using the Kremer-Grest model, which has proven itself for decades as a realistic and robust model in rheological studies of polymer dynamics~\cite{Grest1990}. Using the chain molecules allowed to virtually remove the gas phase in the simulations, to simplify evaluation of the free surface profiles, and to smoothly control the liquid viscosity by gradually varying the chain length $N_B$. The state of the liquid, its temperature $\displaystyle T_0=0.8\,\epsilon_{ \mathit{ff} }/k_B$ ($k_B$ is the Boltzmann constant) was locally controlled by means of a dissipative particle dynamics thermostat with the cut-off distance of $2.5\, \sigma_{\mathit{ff}}$, matching the cut-off distance of the LJ potential, and friction $\varsigma_{dpd}=0.5\tau_0^{-1}$, $\tau_0=\sigma_{\mathit{ff}}\sqrt{\frac{m_f}{\epsilon_{\mathit{ff}}}}$ to have minimal side effects on particle dynamics~\cite{Kremer2003}. The bulk density was $\rho=0.91\,\sigma_{\mathit{ff}}^{-3}$  with no gas phase present in the problem. The droplet depth in the periodic direction was set to $\Delta x=18\,\sigma_{\mathit{ff}}$  to be short enough to suppress the Plateau-Rayleigh instability. The solid wall particles were attached to anchor points forming the lattice layers {\it via} harmonic potential chosen such that to neglect elasto-capillarity effects~\cite{Weijs2013}. The anchor points in the layer of the solid wall facing the liquid molecules have been slightly randomized in the vertical $y$ direction, with the amplitude $\sqrt{\langle \delta y^2\rangle}=0.3\,\sigma_{\mathit{ff}}$, to avoid any bias towards ideal substrates in the study. The solid wall particles were moving with velocity $U$ in the $z$-direction, where a reflective wall acted as a barrier to simulate a forced wetting regime in a moving meniscus. The substrate particle density in most simulations (unless otherwise stated) was set to $\Pi_S=4.1\,\sigma_{\mathit{ff}}^{-3}$ with $\sigma_{\mathit{wf}}=0.7\,\sigma_{\mathit{ff}}$ and $\epsilon_{\mathit{wf}}=0.9\,\epsilon_{\mathit{ff}}$ to have a static contact angle $\theta_0=39\pm 3^{\circ}$. 

\begin{figure}
\begin{center}
\includegraphics[trim=0cm 0.5cm 0cm 0cm,width=\columnwidth]{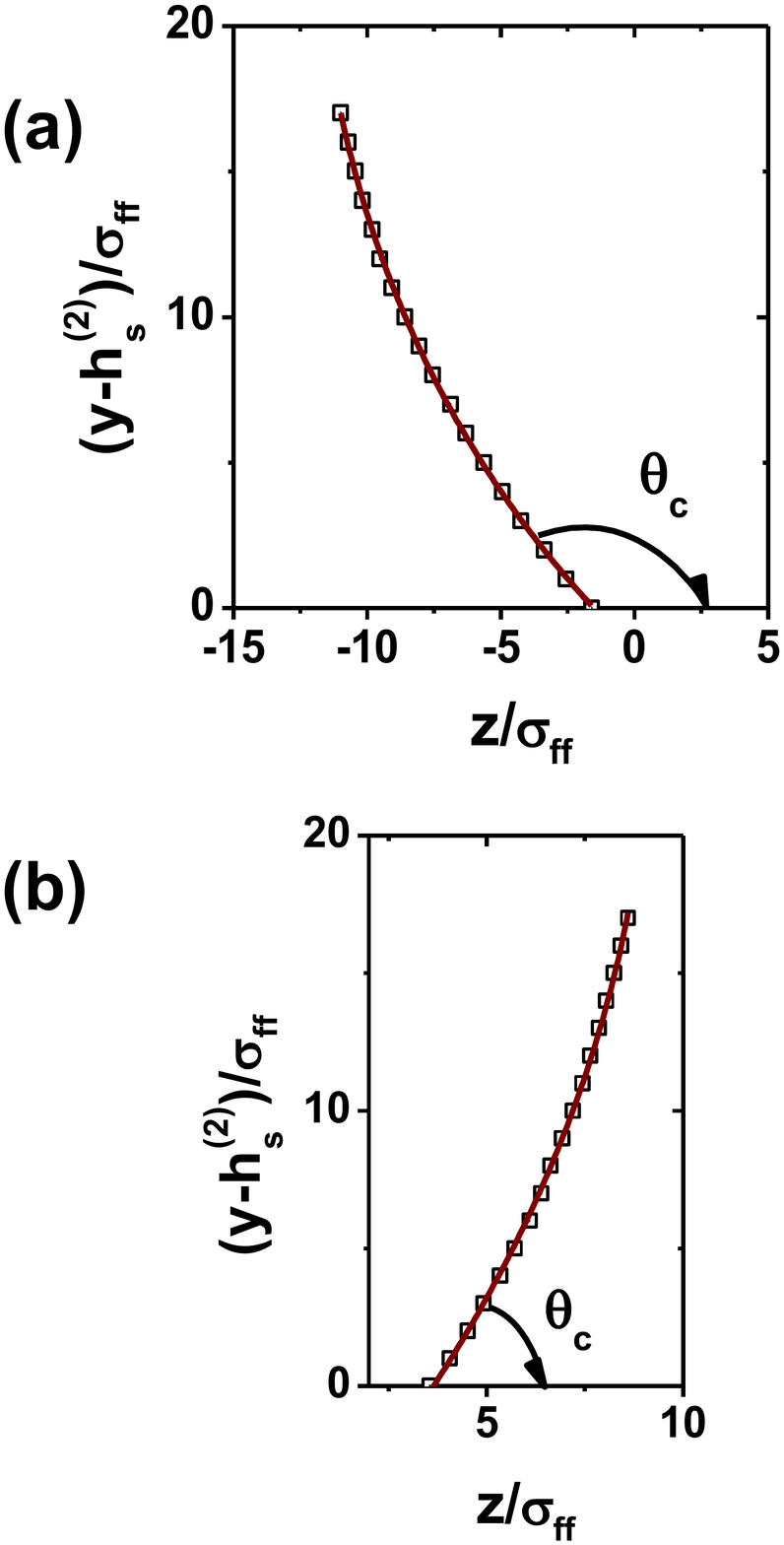}
\end{center}
\caption{Illustration of the free surface profiles (equimolar surfaces) developed from the MDS at $N_B=5$, $\displaystyle T_0=0.8\,\epsilon_{ \mathit{ff} }/k_B$ and $\theta_0=39\pm 3^{\circ}$, and the circular fits $(\bar{y}-y_0)^2+(z-z_0)^2=R^2$ to them, where $\bar{y}=y-h_s^{(2)}$ and the distances $y$ and $z$ are measured as in Figure \ref{Fig2a}. Here, (a) $\theta_c=136\pm 5^{\circ}$, $Ca=1.14$, $R=34.8\pm 1.7\, \sigma_{\mathit{ff}}$, $y_0=24.9\pm 1.1\, \sigma_{\mathit{ff}}$, $z_0=23.2\pm 1.6\, \sigma_{\mathit{ff}}$, (b) $\theta_c=65\pm 4^{\circ}$, $Ca=0.057$, $R=59.7\pm 5\, \sigma_{\mathit{ff}}$, $y_0=25.2\pm 1.6\, \sigma_{\mathit{ff}}$, $z_0=-51.7\pm 4.7\, \sigma_{\mathit{ff}}$. } 
\label{Fig-profile}
\end{figure}

After initial equilibration during $10000\,\tau_0$ with the time step $0.01\,\tau_0$, used in the study, we reached a steady state to take measurements of dynamic contact angle, density and velocity distributions in the interfacial layers and in the bulk. Distributions of density and velocity were averaged over the droplet depth $\Delta x$ and a time interval $10000\,\tau_0$, unless otherwise specified. The dynamic contact angle $\theta_c$ was a function of velocity $U$, Figure \ref{Fig2a} and Figure \ref{Fig2b}, determined from the free surface profiles defined as the locus of equimolar points of density distributions,  Figure \ref{Fig-profile}. The location of the equimolar surface has been determined by measuring density distributions along the $z$-axis in the layers of $\Delta y=1\,\sigma_{\mathit{ff}}$ thickness with the spatial resolution $\Delta z\approx 0.3\,\sigma_{\mathit{ff}}$. The size of the observation box along the $z$-direction was chosen to be small enough to resolve the observed liquid-gas interface density profiles, Figures \ref{Fig2a}, \ref{Fig5} and \ref{Fig6}. 

The free surface profiles were developed by means of a three-parameter $(R,y_0,z_0)$ circular fit 
\begin{equation}
\label{Circular}
(\bar{y}-y_0)^2+(z-z_0)^2=R^2,
\end{equation} 
which has been applied to the positions of the equimolar points obtained in the MDS excluding a layer adjacent to the substrate of thickness $h_s^{(2)}=4\,\sigma_{\mathit{ff}}$ corresponding to the liquid-solid interface, Figure \ref{Fig2a} (magnified view). Here, $\bar{y}=y-h_s^{(2)}$ and the distance $y$ is measured as in Figure \ref{Fig2a}. The choice of the fitting function has been dictated by the fitting accuracy, see Figure \ref{Fig-profile}, and to some extent by the fact that at small capillary numbers $Ca\ll 1$ (low viscous stresses), the free surface profile is bound to be circular~\cite{Shikhmurzaev-2007}. Similar fitting functions were used to evaluate free surface profiles of liquids drops in the absence of gravity and at low capillary numbers, when the surface shape is expected to be spherical~\cite{Blake2009}.

At large capillary numbers, $Ca\sim 1$, the free surface shape may not be always circular. So that in general the circular fit (\ref{Circular}) has been only applied to a part of the free-surface profile of length $\approx 20\,\sigma_{\mathit{ff}}$. As one can see, Figure \ref{Fig-profile} (a), even in this case, $Ca=1.14$, the fit has demonstrated very good accuracy. We have verified that changing the arc length by approximately $ \pm 5\,\sigma_{\mathit{ff}}$ at a fixed $h_s^{(2)}$ produced an uncertainty in the contact angle determination not more than $\Delta \theta_c \le 1^{\circ}$.  

The necessity to cut off a layer of molecules at the solid substrate to calculate a contact angle is due to strong bending of the equimolar surface observed at the solid substrate on the length scale of a few atomic distances, Figures \ref{Fig2a}, \ref{Fig5} and \ref{Fig6}. Apparently, this was due to the strong density perturbations, the particle layering, induced by the solid wall potential commonly observed in MDS at the contact lines~\cite{Leroy2010}. The question then arises whether the bended area should be included in the calculations of contact angles or not. We have clarified this issue previously, as other groups did~\cite{Blake2017}, by directly probing the Young-Dupr\'{e} equation in equilibrium conditions by placing a cylindrical (to avoid possible line tension effects) drop of $40 000$ particles on the substrate~\cite{Nano2016}. The static contact angle was then obtained in two ways: first by direct measurements from the free surface profiles using the same methodology as described above and second, for comparison, {\it via} the Young-Dupr\'{e} equation using independently calculated  equilibrium surface tensions. We have found a very good agreement between the two angles, when the highly bend region was excluded from the contact angle evaluation procedure, in full accordance with the fact that the contact angle is an experimentally observed, macroscopic quantity. We also note, that slightly increasing the cut-off distance by $2\,\sigma_{\mathit{ff}}$ produced an uncertainty in the contact angle determination not more than $\Delta \theta_c \le 1^{\circ}$. One needs of course to keep the cut-off distance at a minimum in MDS given usually the small size of the systems. 

 We would like to note that the particular set of MDS parameters used and discussed in this work is representative of a larger set of MDS simulations produced in the previous work~\cite{Nano2016}. The simulations were performed for different model liquids with $N_B$ ranging from $N_B=1$ to $N_B=30$, at different static contact angles $\theta_0$ in the range $0^{\circ} \le \theta_0\le 106^{\circ}$, at different system sizes $H$, Figure \ref{Fig2a} and Figure \ref{Fig2b}, ranging from $H=40\,\sigma_{\mathit{ff}}$ to $H=100\,\sigma_{\mathit{ff}}$, at different contact line velocities $0.005\, u_0 < U \le 0.2 \, u_0$ ($u_0=\sqrt{\epsilon_{\mathit{ff}}/m_{\mathit{f}}}$) and at different temperatures $0.8 \,\epsilon_{ \mathit{ff} }/k_B \le T_0 \le 1 \,\epsilon_{ \mathit{ff} }/k_B$, see details in the previous work\cite{Nano2016}. While the value of the macroscopic model parameters, such as $Re$, $Ca$, $\rho_s^{(1)}$, $\rho_s^{(2)}$ and $L_s$, which we consider in detail in the next section, might differ from one case to another, there was no qualitative differences observed in the flow kinematics discussed in this work and in the other cases studied~\cite{Nano2016}.   

\subsection*{Macroscopic parameters from MDS}
To compare the results of MDS with the macroscopic description  (\ref{Stokes})-(\ref{CLfluxND}), the model parameters $Re$, $Ca$, $\rho_s^{(1)}$, $\rho_s^{(2)}$ and $L_s$ were directly defined from the MDS. The capillary number $Ca=\mu U/\gamma$ was obtained from liquid viscosity $\mu=10.5\,  \sqrt{ \epsilon_{\mathit{ff}} m_{\mathit{f}} } /  \sigma_{\mathit{ff}}^2$ and surface tension $\gamma=0.92\, \epsilon_{ \mathit{ff} } / \sigma_{ \mathit{ff} }^2$, calculated as in reference~\cite{Lukyanov2013-1}. In our approach here, there is some degree of arbitrariness how one can set apart the interfaces from the bulk~\cite{Rowlinson-Widom-1989, Shikhmurzaev-2007}. To get an estimate for characteristic values of interfacial parameters a minimum cut-off the areas with strong variations of density is applied. Density perturbations induced by the solid wall potential fade away at $y \approx 4\,\sigma_{\mathit{ff}}$ counting from the equimolar point of solid wall particles, see the magnified view in Figure \ref{Fig2a}, with a similar length scale observed in the liquid-gas interfaces. The position of interfaces is defined according to $\rho_s=const$. The liquid-solid interfacial shape is then a straight line at a fixed distance from the substrate $h_s^{(2)}=4\,\sigma_{\mathit{ff}}$. Accordingly, the position of the free surface facing the liquid bulk was set as the locus of points equidistant from the equimolar surface of the liquid-gas interface density profiles in the normal to the equimolar surface direction. For consistency, the free surface should cross the liquid-solid interface boundary at a point where no variations of the surface density is observed, $\rho_s^{(2)}=const$, Figure \ref{Fig2a}. This way the width of the liquid-gas interfacial layer and the position of the contact line, at $z_{CL}\approx 4\,\sigma_{ \mathit{ff} }$ in this case, are fully defined. Integrating the particle density distribution across the interfaces, one gets $\rho_s^{(1)}\approx 5.5\,\sigma_{ \mathit{ff} }^{-2}$ and $\rho_s^{(2)}\approx 3.5\,\sigma_{ \mathit{ff} }^{-2}$. 

The apparent slip length $L_s=\frac{\mu}{\beta}$ can be obtained from MDS in homogeneous film flow conditions by measuring velocity profiles $u_z(\bar{y})$ at $\bar{y}=0$, $\bar{y}=y-h_s^{(2)}$. From (\ref{NSOr}) in the absence of gradients in the $z$-direction, $L_s\frac{\partial u_z}{\partial \bar{y}}=u_z$ and we have obtained $L_s=3.8\pm 0.8\, \sigma_{ \mathit{ff} }$.  

\begin{figure}
\begin{center}
\includegraphics[trim=0cm 2.5cm 2cm 0cm,width=\columnwidth]{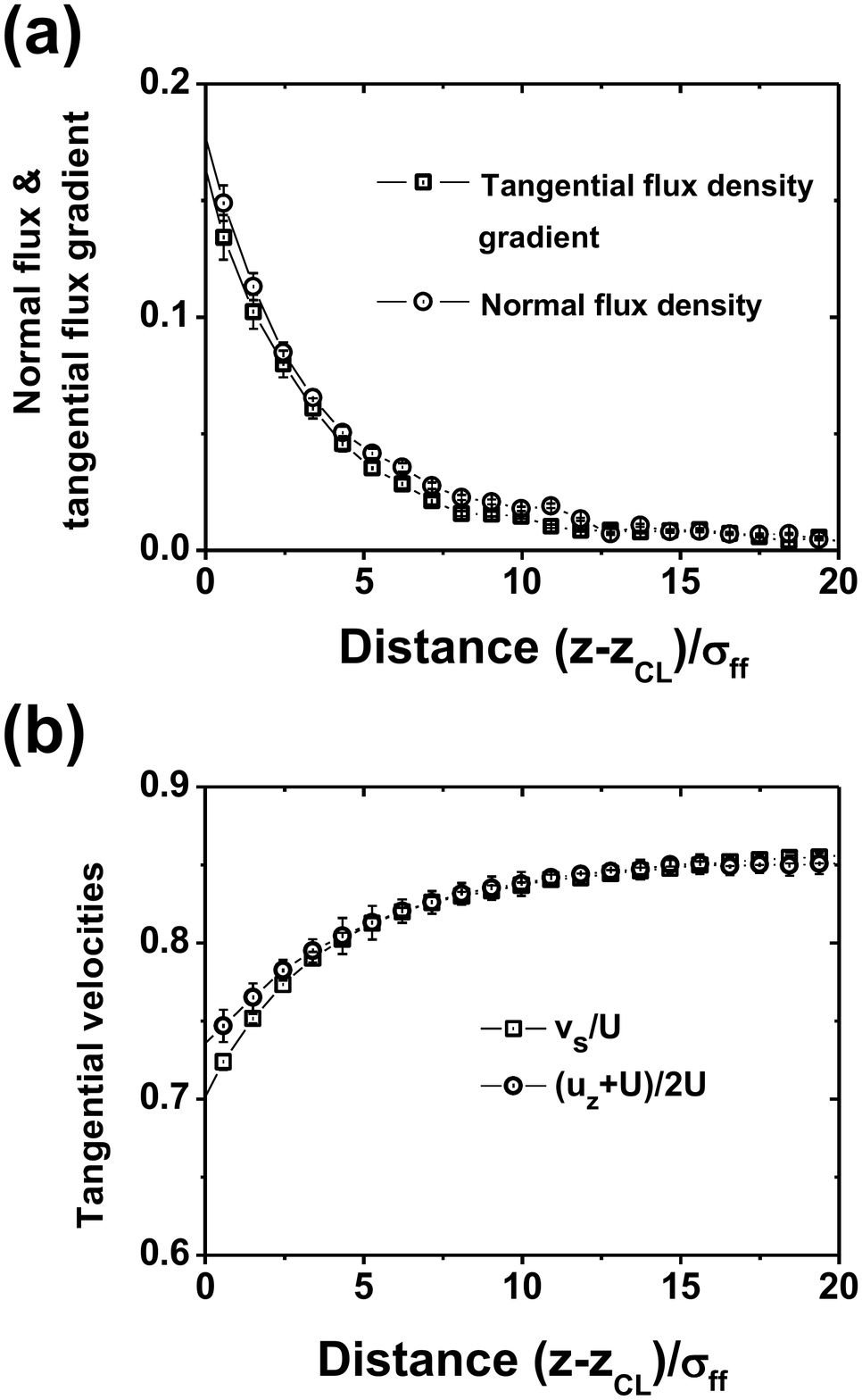}
\end{center}
\caption{MDS results at $U=0.1\,u_0$, $Ca=1.14$, $\theta_c=136\pm5^{\circ}$, $u_0=\sqrt{\epsilon_{\mathit{ff}}/m_{\mathit{f}}}$: (a) Distributions of the reduced normal flux density $\dfrac{u_n}{U}$ and the reduced tangential flux density gradient in the $z$-direction $\dfrac{1}{\rho U}\dfrac{\partial (\rho_s {v}_s)}{\partial z}$ at the liquid-solid interface of the moving droplet, Figure \ref{Fig2b}, at $h_s^{(2)}=4\,\sigma_{\mathit{ff}}$ to verify conservation of mass in the boundary layer (\ref{MImpC}). (b) Distributions of the reduced surface velocity $\dfrac{v_s}{U}$ and $\dfrac{u_z +U}{2U}$ to verify boundary condition (\ref{Coutte}) at the liquid-solid interface in the MDS at $h_s^{(2)}=4\,\sigma_{\mathit{ff}}$.  In the plots, distance $z$ is measured as in Figure \ref{Fig2a}. The MDS results were averaged over five statistically independent simulations.} 
\label{Fig3}
\end{figure}

\begin{figure}
\begin{center}
\includegraphics[trim=0cm 2.5cm 2cm 0cm,width=\columnwidth]{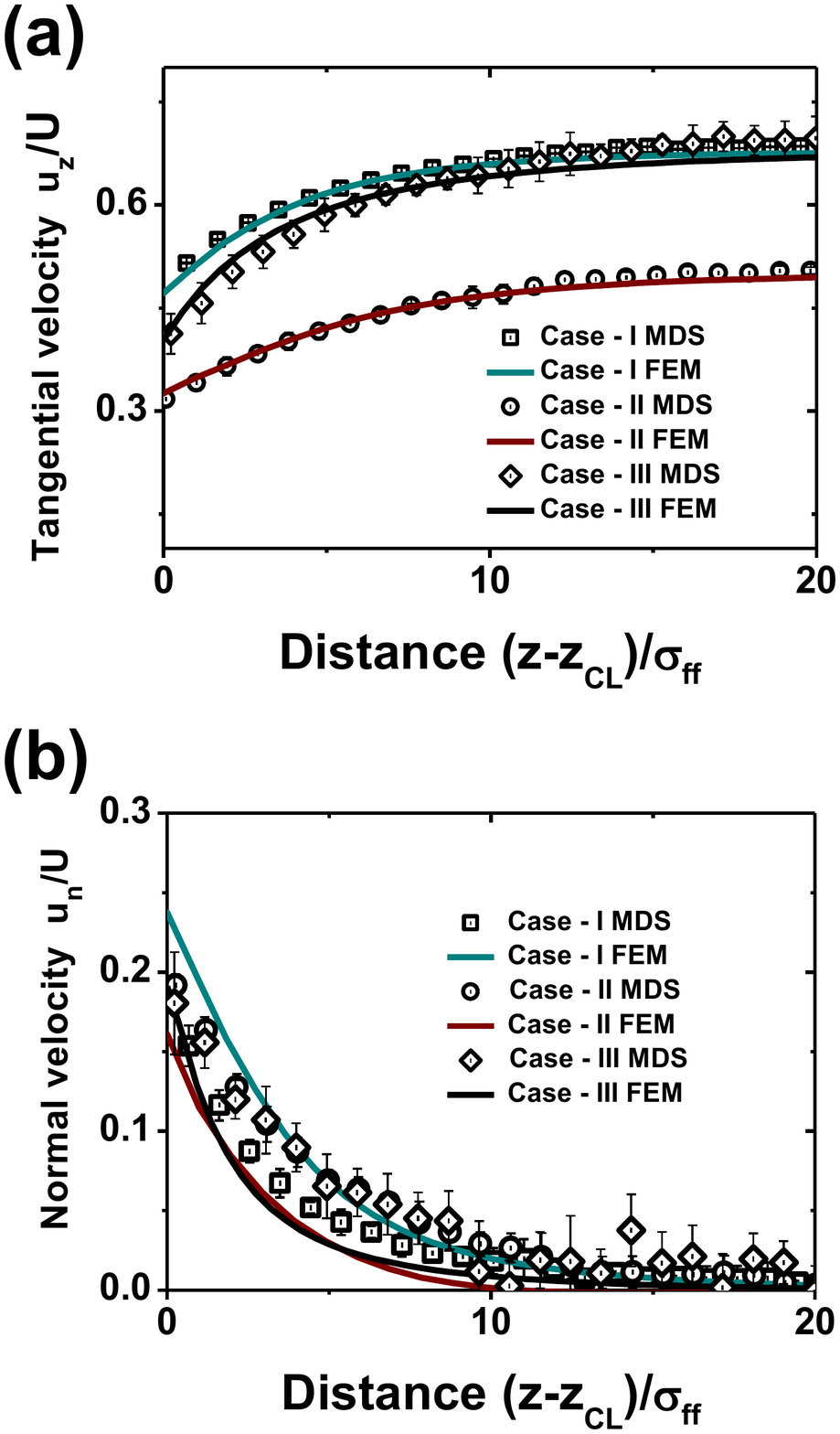}
\end{center}
\caption{A comparison between continuum simulations and MDS in three cases: {\bf I} at $U=0.1\,u_0$, $Ca=1.14$, $\theta_c=136\pm5^{\circ}$, $h_s^{(2)}=4\,\sigma_{\mathit{ff}}$, $\alpha_1=1.6$, $\alpha_2=0.51$ and $L_s=3.8\pm 0.8\, \sigma_{ \mathit{ff} }$, {\bf II} at $U=0.1\,u_0$, $Ca=1.14$, $\theta_c=136\pm5^{\circ}$, $h_s^{(2)}=6\,\sigma_{\mathit{ff}}$, $\alpha_1=0.93$, $\alpha_2=0.36$ and $L_s=8\pm 0.9\, \sigma_{ \mathit{ff} }$, {\bf III} at $U=0.03\,u_0$, $Ca=0.34$, $\theta_c=114\pm5^{\circ}$, $h_s^{(2)}=4\,\sigma_{\mathit{ff}}$, $\alpha_1=1.6$, $\alpha_2=0.51$ and $L_s=3.8\pm 0.8\, \sigma_{ \mathit{ff} }$. Here, parameter $u_0=\sqrt{\epsilon_{\mathit{ff}}/m_{\mathit{f}}}$. (a) Distributions of the tangential velocity $u_{z}$ at the liquid-solid interface. (b) Distributions of the normal velocity $u_{n}$ at the liquid-solid interface. The solid line is continuum finite-element simulations, symbols represent the results of MDS. In the plots, distance $z$ is measured as in Figure \ref{Fig2a}. The MDS results were averaged over five statistically independent simulations.} 
\label{Fig3a}
\end{figure}

\section*{Results and Discussion}
What do we observe in the MDS? In the bulk of the liquid domain, the flow demonstrates clear rolling motion with non-zero flux in and out of both interfacial layers at the contact line, as is qualitatively expected from the modified boundary conditions (\ref{MImpC}) and (\ref{MImpCf}), where the impermeability condition has been relaxed, Figure \ref{Fig2b}, the magnified view. This behaviour is typical, irrespective of the contact angle (obtuse or acute), and it has been also observed in MDS using a different set-up~\cite{Blake2015}. The tangential and normal to the substrate velocities $u_{z}$, $v_s$ and $u_n$ are clearly non-zero at the contact line, demonstrating no stagnation zone or any obstacle to the flow both in the bulk and in the interfaces, Figure \ref{Fig3}(a)-(b). Note, that the velocities $u_z$ and $u_n$ have been measured in MDS via averaging over a layer parallel to the substrate of thickness $\Delta y=2\,\sigma_{\mathit{ff}}$ centered at $y=h_s^{(2)}$. At the same time, the surface velocity $v_s$ has been obtained via averaging in the whole layer adjacent to the substrate of thickness $\Delta y=h_s^{(2)}$. In all velocity distribution measurements spatial resolution in the $z$-direction was $\Delta z \approx 1\,\sigma_{\mathit{ff}}$. 

In our parameter range $Re=\frac{\rho U L_s}{\mu}\ll 1$, and the flow is in the Stokes regime. At the same time the length scales involved are in the mesoscopic range, that is $L_s\approx h_s^{(2)}$. The question is can we simulate this kind of flows using macroscopic description  (\ref{Stokes})-(\ref{CLfluxND}) and with what accuracy?  

Obviously, conditions (\ref{MImpCND}), (\ref{MImpCfND}) and (\ref{CLfluxND}) representing conservation of mass should be always fulfilled if conditions (\ref{Free-surface-no-slip}) and (\ref{Coutte}) are satisfied, which is the case, as one can see from Figure \ref{Fig3}(b). The conservation of mass principle itself is satisfied in the MDS with high accuracy, Figure \ref{Fig3}(a). The other boundary conditions would be difficult to validate directly since the notion of stress tensor is not well defined on mesoscopic length scales~\cite{Schofield1982, Rowlinson-Widom-1989}. 

\begin{figure}
\begin{center}
\includegraphics[trim=0cm 1.5cm 1cm 0cm,width=1\columnwidth,angle=0]{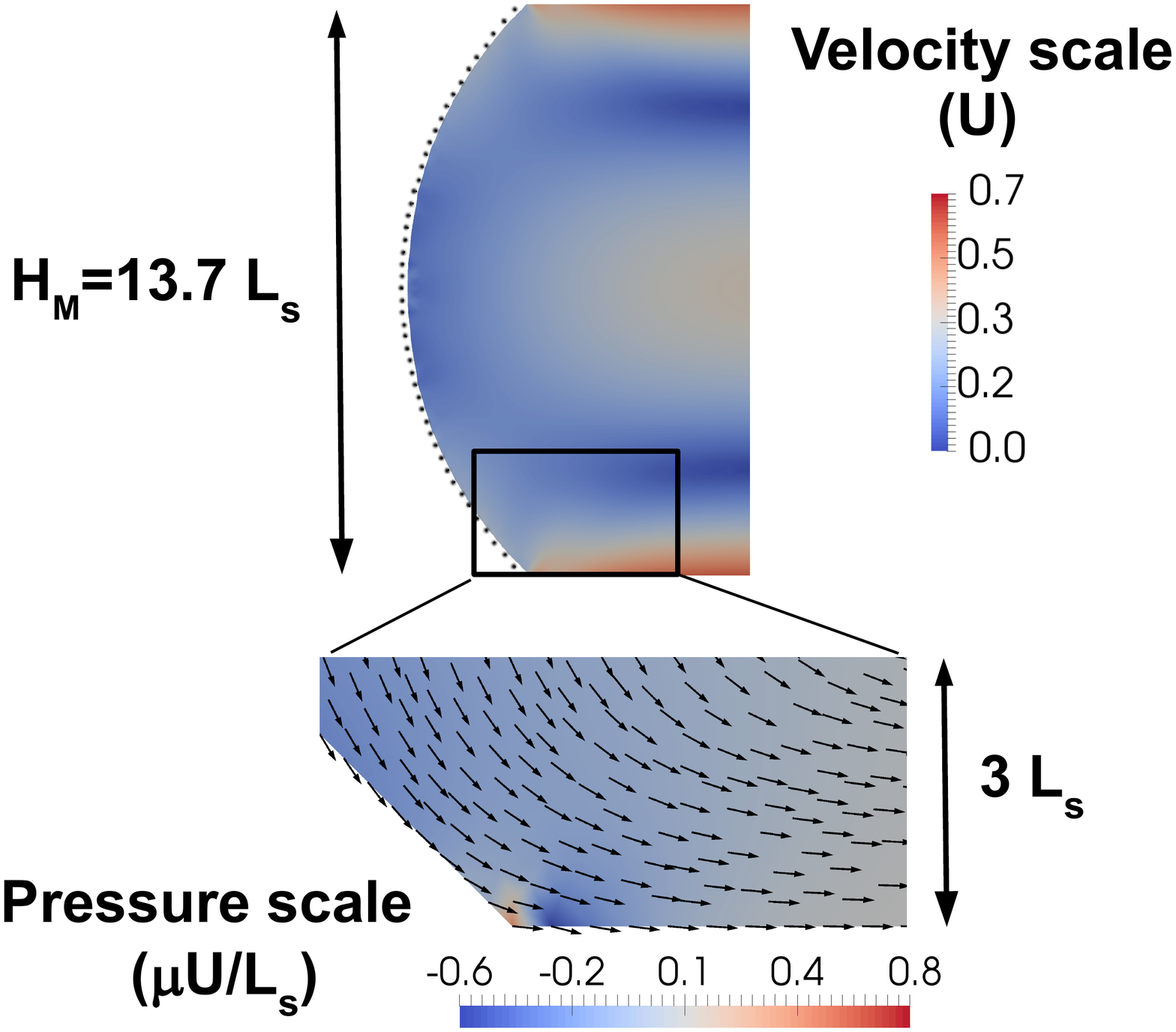}
\end{center}
\caption{A sample of typical continuum simulations with a dynamic contact angle of $\theta_c = 136^\circ$ at $Ca=1.14$. The macroscopic parameters are $\alpha_1=1.6$, $\alpha_2=0.51$ and $L_s=3.8\pm 0.8\, \sigma_{ \mathit{ff} }$. The top plot shows the final profile after an ALE finite element scheme is applied to the problem. Notice the continuum simulation is a good approximation of the MD simulations which are overlaid as black dots on the profile. The colour indicates velocity amplitude. The bottom plot shows the velocity profile near the contact angle and distribution of pressure (the colour map).} 
\label{Fig4b}
\end{figure}

\begin{figure}
\begin{center}
\includegraphics[trim=0cm 1.5cm 1cm 0cm,width=1\columnwidth,angle=0]{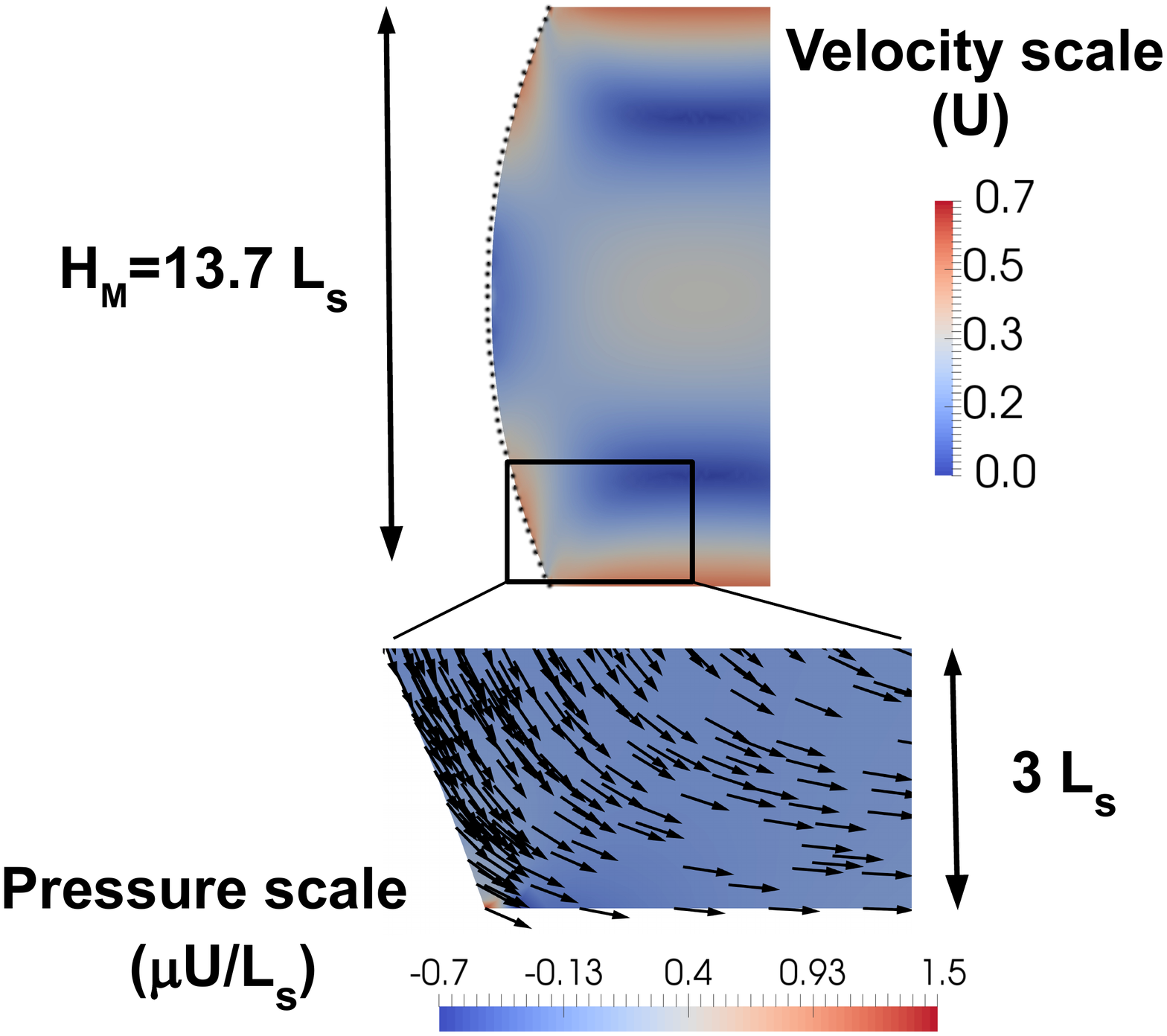}
\end{center}
\caption{A sample of typical continuum simulations, similar to Figure \ref{Fig4b}. Here it is with a dynamic contact angle of $\theta_c = 114^\circ$ at $Ca=0.34$ and a set of the macroscopic parameters $\alpha_1=1.6$, $\alpha_2=0.51$ and $L_s=3.8\pm 0.8\, \sigma_{ \mathit{ff} }$.} 
\label{Fig4a}
\end{figure}

\begin{figure}
\begin{center}
\includegraphics[trim=0cm 1.5cm 1cm 0cm,width=1\columnwidth,angle=0]{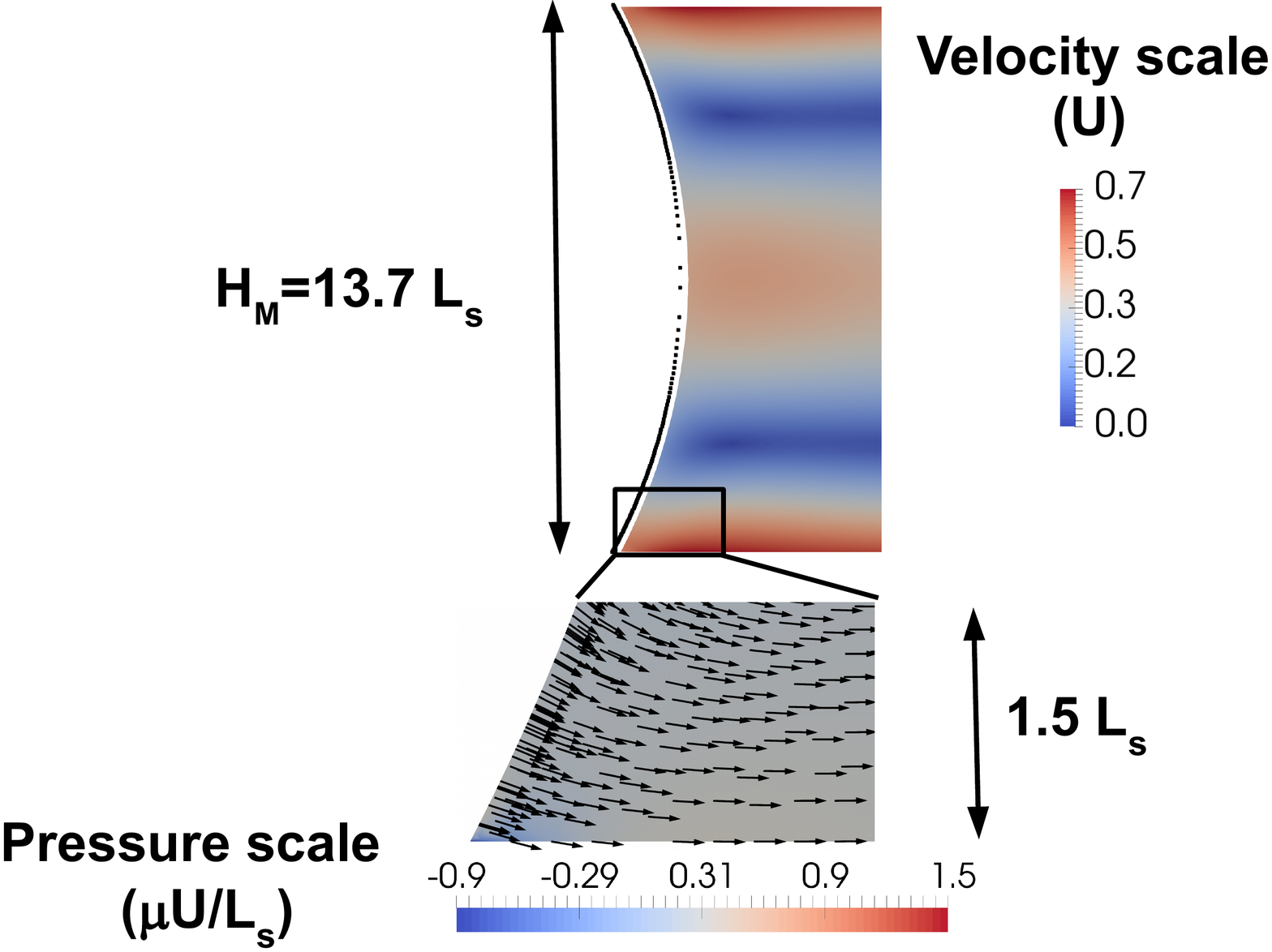}
\end{center}
\caption{A sample of typical continuum simulations, similar to Figure \ref{Fig4b}. Here it is with a dynamic contact angle of $\theta_c = 65^\circ$ at $Ca=0.057$ and a set of the macroscopic parameters $\alpha_1=1.6$, $\alpha_2=0.51$ and $L_s=3.8\pm 0.8\, \sigma_{ \mathit{ff} }$.} 
\label{Fig4d}
\end{figure}


To verify the accuracy of the macroscopic approach, we use an arbitrary Lagrangian-Eulerian (ALE) finite-element simulation of the problem (\ref{Stokes})-(\ref{CLfluxND}) with the parameters and geometry taken directly from the MDS set-up. That is, a liquid flow is simulated in steady conditions in a macroscopic droplet forced to move in between two solid substrates separated by the distance $H_M=13.7\, L_s$ using the macroscopic set of parameters $\alpha_1=1.6$ and $\alpha_2=0.51$ directly calculated from the set of interfacial parameters $h_s^{(2)}=4\,\sigma_{\mathit{ff}}$, $L_s=3.8\,\sigma_{\mathit{ff}}$, $\rho_s^{(1)} = 5.5\,\sigma_{ \mathit{ff} }^{-2}$ and $\rho_s^{(2)} = 3.5\,\sigma_{ \mathit{ff} }^{-2}$ obtained in the MDS. The macroscopic parameter $H_M$ was obtained by deducting the width of the two liquid-solid interfacial layers $2h_s^{(2)}=8\,\sigma_{\mathit{ff}}$ from the MDS set-up width $H=60\,\sigma_{\mathit{ff}}$. 

\begin{figure}
\begin{center}
\includegraphics[trim=0cm 1.5cm 1cm 0cm,width=\columnwidth,angle=0]{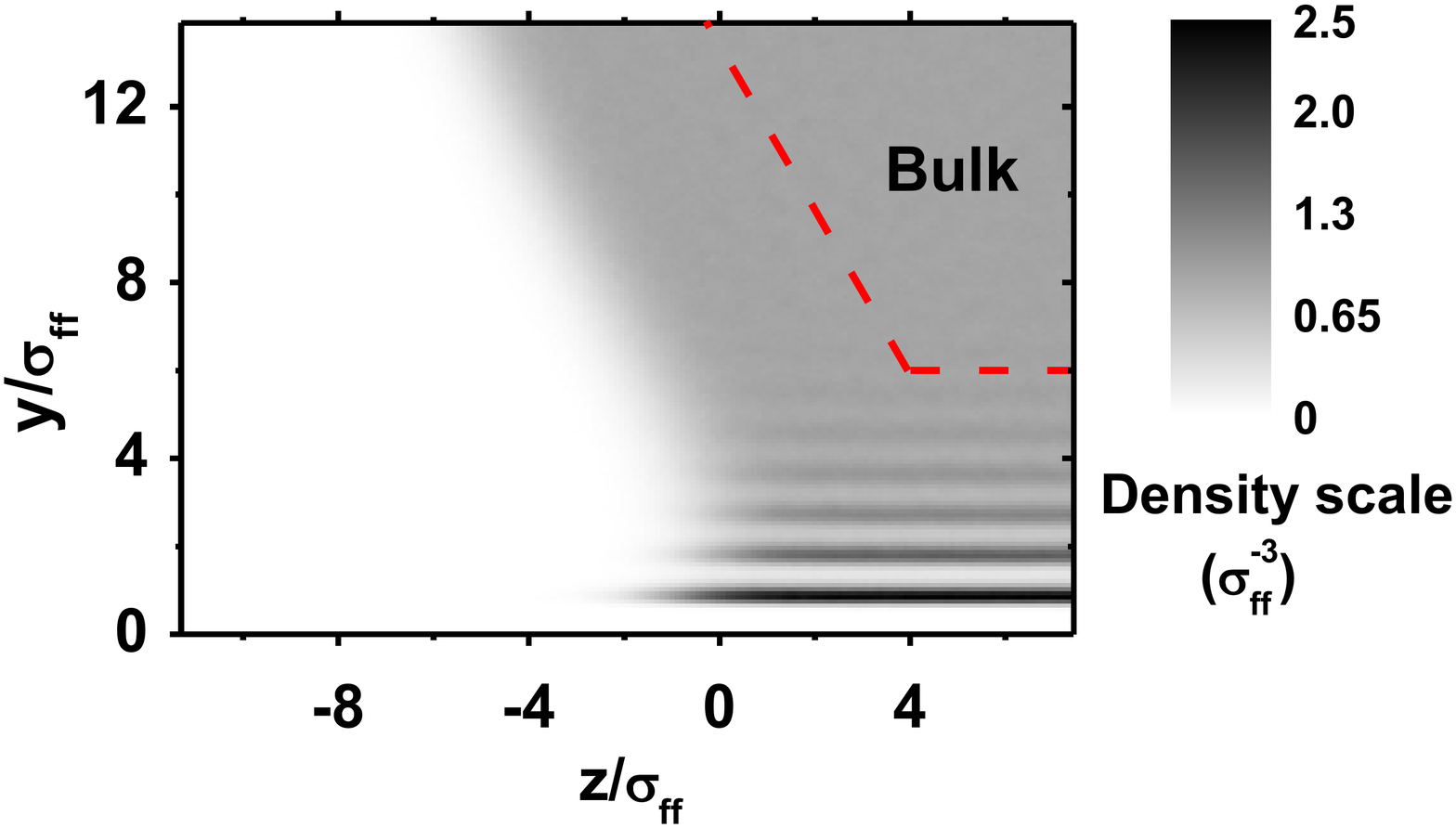}
\end{center}
\caption{Distribution of the particle density at the contact line of a moving droplet at $U=0.06\, u_0$ ($Ca\approx 0.69$) and a dynamic contact angle $\theta_c=123\pm 5^{\circ}$, $u_0=\sqrt{\epsilon_{\mathit{ff}}/m_{\mathit{f}}}$. The static contact angle $\theta_0=0^{\circ}$ at $\epsilon_{\mathit{wf}}=1.2\,\epsilon_{\mathit{ff}}$ and $\sigma_{\mathit{wf}}=\,\sigma_{\mathit{ff}}$. The dashed lines (red) designate the bulk-interface boundary. In the plot, distances $y$ and $z$ are measured as in Figure \ref{Fig2a}.} 
\label{Fig5}
\end{figure}

The dynamic contact angle $\theta_c$ was fixed to the values observed in MDS at a particular substrate velocity $U$. The results of simulations at different velocities $U$ and dynamic contact angles, in particular at $\theta_c=136^{\circ}$, $Ca=1.14$, $Re=0.033$ (as in Figure \ref{Fig2b}), at $\theta_c=114^{\circ}$, $Ca=0.34$, $Re=0.01$ (as in Figure \ref{Fig2a}) and at $\theta_c = 65^\circ$, $Ca=0.057$, $Re=1.65\times 10^{-3}$, are shown in Figures \ref{Fig3a}(a)-(b), \ref{Fig4b}, \ref{Fig4a} and \ref{Fig4d}. One can see that continuum simulations can correctly reproduce all qualitative features of the global velocity field in the bulk, see Figures  \ref{Fig4b}, \ref{Fig4a} and  \ref{Fig4d}.  The results of continuum simulations are also in a very good quantitative agreement with the MDS results. Consider, for example, distributions of tangential $u_z$ and normal $u_n$ velocities at the solid substrate, Figure \ref{Fig3a}(a)-(b), and the shape of the free surface profiles, Figures \ref{Fig4b}, \ref{Fig4a} and \ref{Fig4d}. Note, the pressure in the continuum solution to the problem (\ref{Stokes})-(\ref{CLfluxND}) is regular, Figures  \ref{Fig4b}, \ref{Fig4a} and \ref{Fig4d}, and the free surface profiles have no concave bending at obtuse contact angles typical with a logarithmic singularity of the pressure field. This was exactly observed in the MDS, Figure \ref{Fig2a} and Figure \ref{Fig2b}, and Figures \ref{Fig4b}, \ref{Fig4a} and \ref{Fig4d}, and in the nanoscale experiments~\cite{Chen2014}. The characteristic value of the pressure at the contact line obtained in the FEM simulations is consistent with the viscous stresses, which are expected to be developed on the length scale of one slip length $L_s$, that is $p\approx \mu U/L_s$. Note, that the pressure regularity also suggests, that in our case there are no spurious solutions reported in reference~\cite{Sprittles2011}. 

To probe the sensitivity of the methodology to the cut-off distance $h_s^{(2)}$, we have performed FEM simulations with a different set of macroscopic parameters $\alpha_1=0.93$ and $\alpha_2=0.36$ corresponding to the set of interfacial parameters obtained in the MDS using a different value of the cut-off distance $h_s^{(2)}=6\,\sigma_{\mathit{ff}}$, that is subsequently  $L_s=8\pm 0.9\, \sigma_{ \mathit{ff} }$, $\rho_s^{(1)} = 6.8\,\sigma_{ \mathit{ff} }^{-2}$ and $\rho_s^{(2)} = 5.3\,\sigma_{ \mathit{ff} }^{-2}$. The width of the macroscopic set-up was then set to $H_M=6\, L_s$. As is seen from Figure \ref{Fig3a}(a)-(b), a similar approximation can be achieved by the continuum model. So, that the methodology is practically invariant to the cut-off procedure.  

How sensitive are the macroscopic parameters of the model to the properties of the solid substrate, such as the substrate density $\Pi_S$, the strength $\epsilon_{\mathit{wf}}$ and the length scale $\sigma_{\mathit{wf}}$ of the interaction potential? Apparently those parameters affect the state of the liquid-solid interface, its surface tension and, as a consequence, the wettability of the solid surface, that is the static contact angle $\theta_0$. 

Consider a droplet of the same liquid with the number of beads $N_B=5$ at $\displaystyle T_0=0.8\,\epsilon_{ \mathit{ff} }/k_B$ moving in the same geometry over a substrate with the particle density set to $\Pi_S=1.4\,\sigma_{\mathit{ff}}^{-3}$ and with the interaction potential $\epsilon_{\mathit{wf}}=1.2\,\epsilon_{\mathit{ff}}$ with the length scale $\sigma_{\mathit{wf}}=\,\sigma_{\mathit{ff}}$ to have a static contact angle $\theta_0=0^{\circ}$. The distribution of the particle density in this case is shown in Figure \ref{Fig5} at $Ca\approx 0.69$. One can readily observe that due to the larger length scale $\sigma_{\mathit{wf}}=\,\sigma_{\mathit{ff}}$ and essentially larger characteristic interaction energy $\epsilon_{\mathit{wf}}$, the density perturbations extend deeper into the liquid volume. The minimal cut-off distance turned out to be $h_s^{(2)}=6\,\sigma_{\mathit{ff}}$ in this case. 

\begin{figure}
\begin{center}
\includegraphics[trim=0cm 1.5cm 1cm 0cm,width=1\columnwidth,angle=0]{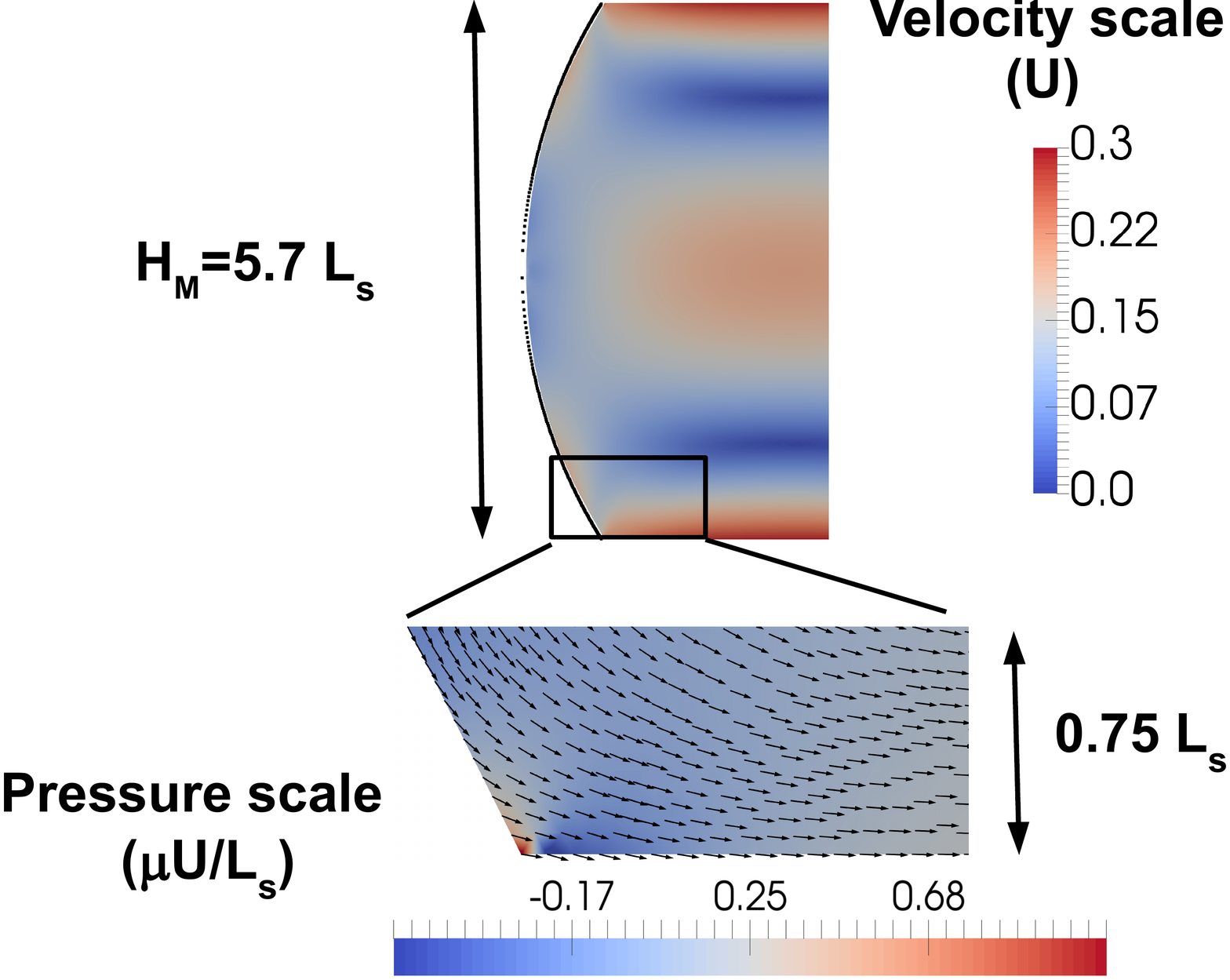}
\end{center}
\caption{A sample of typical continuum simulations, similar to Figure \ref{Fig4b}. Here it is with a dynamic contact angle of $\theta_c = 123^\circ$ at $Ca=0.69$, the static contact angle $\theta_0 = 0^\circ$ and a set of the macroscopic parameters $\alpha_1=0.62$, $\alpha_2=0.33$ and $L_s=8.5\pm 0.9\, \sigma_{ \mathit{ff} }$.} 
\label{Fig4c}
\end{figure}

If we follow the same procedure used in this work to set apart the bulk of the liquid from its interfaces, as is shown in Figure \ref{Fig5}, we obtain that $\rho_s^{(1)} = 4.8\,\sigma_{ \mathit{ff} }^{-2}$ and $\rho_s^{(2)} = 5.1\,\sigma_{ \mathit{ff} }^{-2}$. These values are not far from $\rho_s^{(1)} = 6.8\,\sigma_{ \mathit{ff} }^{-2}$ and $\rho_s^{(2)} = 5.3\,\sigma_{ \mathit{ff} }^{-2}$ obtained for the same cut-off distance, but in the partially wetting case at a static contact angle $\theta_0=39\pm 3^{\circ}$. The apparent slip length $L_s=8.5\pm 0.9\, \sigma_{ \mathit{ff} }$ obtained from MDS in homogeneous film flow conditions by measuring velocity profiles $u_z(\bar{y})$ at $\bar{y}=0$, $\bar{y}=y-h_s^{(2)}$, is practically identical to that observed in the partially wetting case $L_s=8\pm 0.9\, \sigma_{ \mathit{ff} }$. This is no coincidence, since the slip length is apparent and is mostly conditioned by the width of the interface, that is by the cut-off distance $h_s^{(2)}$, and by liquid viscosity. The macroscopic parameters of the model then are $\alpha_1=0.62$ and $\alpha_2=0.33$. A comparison between ALE finite element simulations and the MDS results for this set of parameters is shown in Figure \ref{Fig4c} demonstrating again a very good agreement. 

\begin{figure}
\begin{center}
\includegraphics[trim=0cm 1.5cm 1cm 0cm,width=1\columnwidth,angle=0]{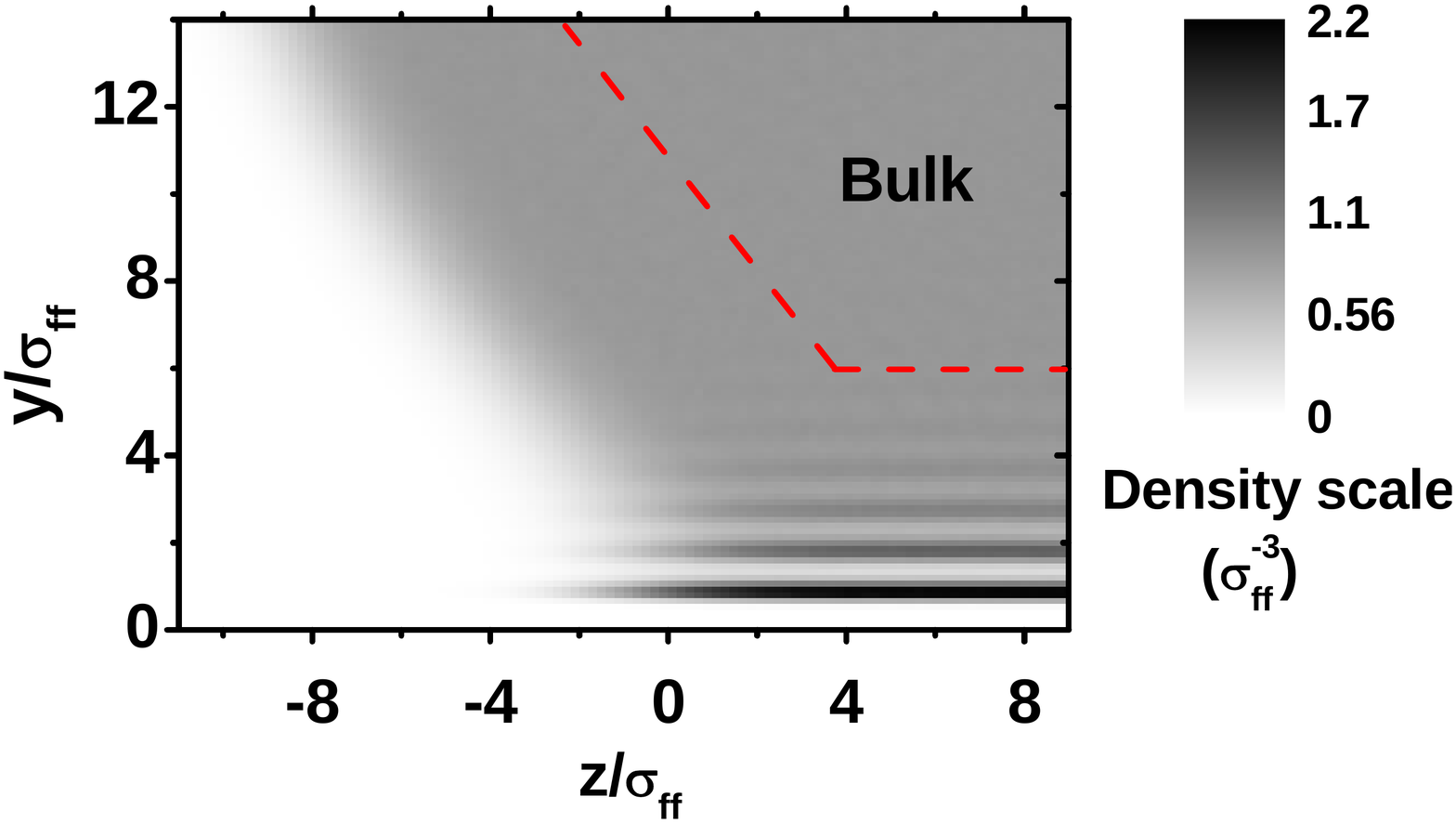}
\end{center}
\caption{Distribution of the particle density at the contact line of a moving droplet at $U=0.1\, u_0$ ($Ca\approx 6.7$) and a dynamic contact angle $\theta_c=137\pm 5^{\circ}$, $u_0=\sqrt{\epsilon_{\mathit{ff}}/m_{\mathit{f}}}$. The static contact angle $\theta_0=0^{\circ}$ at $\epsilon_{\mathit{wf}}=1.3\,\epsilon_{\mathit{ff}}$ and $\sigma_{\mathit{wf}}=\,\sigma_{\mathit{ff}}$. The liquid consists of long-chain molecules $N_B=50$ having macroscopic parameters $\mu=61.8\,  \sqrt{ \epsilon_{\mathit{ff}} m_{\mathit{f}} } /  \sigma_{\mathit{ff}}^2$, $\gamma=0.92\, \epsilon_{ \mathit{ff} } / \sigma_{ \mathit{ff} }^2$ and $\rho=0.89\,\sigma_{\mathit{ff}}^{-3}$ at $\displaystyle T_0=1\,\epsilon_{ \mathit{ff} }/k_B$. The dashed lines (red) designate the bulk-interface boundary. In the plot, distances $y$ and $z$ are measured as in Figure \ref{Fig2a}.} 
\label{Fig6}
\end{figure}

If we change liquid properties, a similar range of parameters is observed. For example, for a liquid consisting of long-chain LJ molecules with $N_B=50$, $\mu=61.8\,  \sqrt{ \epsilon_{\mathit{ff}} m_{\mathit{f}} } /  \sigma_{\mathit{ff}}^2$, $\gamma=0.92\, \epsilon_{ \mathit{ff} } / \sigma_{ \mathit{ff} }^2$ and $\rho=0.89\,\sigma_{\mathit{ff}}^{-3}$ at $\displaystyle T_0=1\,\epsilon_{ \mathit{ff} }/k_B$, using the same procedure, as is shown in Figure \ref{Fig6}, one obtains $\rho_s^{(1)} = 7.5\,\sigma_{ \mathit{ff} }^{-2}$, $\rho_s^{(2)} = 5.1\,\sigma_{ \mathit{ff} }^{-2}$ and $L_s=13\pm 1.1\, \sigma_{ \mathit{ff} }$. 

So one can see that changing the microscopic parameters of the model has no dramatic effect on the values of the macroscopic interfacial parameters.

\section*{Conclusions}
In conclusion, we have demonstrated by comparison with MDS that a simple modification of macroscopic boundary conditions to the Navier-Stokes equations can completely regularize the moving contact-line problem and remove the deficiency of the macroscopic approach with a slip boundary condition. The modification is to relax the impermeability condition and to allow for the mass exchange between interfaces and the bulk area. This is shown to be a key to reproduce the rolling motion observed in MDS and in the simulations using the diffuse interface approaches~\cite{Qian2003}.  In the latter, the impermeability condition is naturally relaxed and a good agreement with MDS was also observed despite the need to involve microscopic length scales to resolve the structure of the interfacial layers. At the same time, our results imply that liquid slippage alone, even with spatially varying slip lengths~\cite{Kirkinis2013}, is insufficient to reproduce the flow kinematics at contact lines. The modified macroscopic model can be used at the nanoscale without producing non-physical effects.

\end{document}